\documentclass[twocolumn,superscriptaddress,
amsmath,amssymb,
aps,
prl,
floatfix,
]{revtex4-1}

\usepackage{miller}

\usepackage{graphicx}                             

\usepackage{dcolumn}
\usepackage{bm} 
\usepackage{color}
\usepackage{siunitx} 
\usepackage{epsfig}
\usepackage{epstopdf}
\usepackage{miller}
\usepackage{ushort}
\ushortCreate:\overline{oshort}

\newcommand{\SIA}{\SI[parse-numbers = false]}

\newcommand{\fcs}{Fe$_{1-x}$Co$_{x}$Si}
\newcommand{\cso}{Cu$_{2}$OSeO$_{3}$}

\begin{document}

\preprint{APS/123-QED}

\title{Reciprocal space mapping of magnetic order in thick epitaxial MnSi films}

\author{B. Wiedemann}
\address{Physik Department, Technische Universit\"at M\"unchen, James-Franck-Strasse 1, 85748 Garching, Germany}

\author{A. Chacon}
\address{Physik Department, Technische Universit\"at M\"unchen, James-Franck-Strasse 1, 85748 Garching, Germany}

\author{S. L. Zhang}
\address{Clarendon Laboratory, Department of Physics, University of Oxford, Parks Road, Oxford, OX1~3PU, UK}

\author{Y. Khaydukov} 
\address{Max-Planck-Institut f\"ur Festk\"orperforschung, Heisenbergstrasse 1, 70569 Stuttgart, Germany}
\address{Max Planck Society, Outstation at FRM-II, D-85748, Garching, Germany}

\author{T. Hesjedal}
\address{Clarendon Laboratory, Department of Physics, University of Oxford, Parks Road, Oxford, OX1~3PU, UK}

\author{O. Soltwedel} 
\address{Physik Department, Technische Universit\"at M\"unchen, James-Franck-Strasse 1, 85748 Garching, Germany}
\address{Max-Planck-Institut f\"ur Festk\"orperforschung, Heisenbergstrasse 1, 70569 Stuttgart, Germany}
\address{Max Planck Society, Outstation at FRM-II, D-85748, Garching, Germany}

\author{T. Keller} 
\address{Max-Planck-Institut f\"ur Festk\"orperforschung, Heisenbergstrasse 1, 70569 Stuttgart, Germany}
\address{Max Planck Society, Outstation at FRM-II, D-85748, Garching, Germany}

\author{S. M\"uhlbauer} 
\address{Forschungsneutronenquelle Heinz Maier Leibnitz (FRMII), Technische Universit\"at M\"unchen, 85748 Garching, Germany}

\author{T. Adams}
\address{Physik Department, Technische Universit\"at M\"unchen, James-Franck-Strasse 1, 85748 Garching, Germany}

\author{M. Halder}
\address{Physik Department, Technische Universit\"at M\"unchen, James-Franck-Strasse 1, 85748 Garching, Germany}

\author{C. Pfleiderer}
\address{Physik Department, Technische Universit\"at M\"unchen, James-Franck-Strasse 1, 85748 Garching, Germany}

\author{P. B\"oni}
\address{Physik Department, Technische Universit\"at M\"unchen, James-Franck-Strasse 1, 85748 Garching, Germany}

\date{\today}

\begin{abstract}
We report grazing incidence small angle neutron scattering (GISANS) and complementary off-specular neutron reflectometry (OSR) of the magnetic order in a single-crystalline epitaxial MnSi film on Si(111) in the thick film limit. Providing a means of direct reciprocal space mapping, GISANS and OSR reveal a magnetic modulation perpendicular to the films under magnetic fields parallel and perpendicular to the film, where additional polarized neutron reflectometry (PNR) and magnetization measurements are in excellent agreement with the literature. Regardless of field orientation, our data does not suggest the presence of more complex spin textures, notably the formation of skyrmions. This observation establishes a distinct difference with bulk samples of MnSi of similar thickness under perpendicular field, in which a skyrmion lattice dominates the phase diagram. Extended x-ray absorption fine structure  measurements suggest that small shifts of the Si positions within the unstrained unit cell control the magnetic state, representing the main difference between the films and thin bulk samples.
\end{abstract}

\pacs{Valid PACS appear here}
\maketitle
The discovery of skyrmions in chiral magnets  \cite{muhlbauer:S:09,neubauer:PRL:09,munzer:PRB:10,jonietz:S:10,yu:N:10,yu:NM:11,seki:S:12,adams:PRL:12,schulz:NP:12,nagaosa:NN:13,milde:S:13,mochizuki:NM:14,schwarze:NM:15} has generated great efforts to exploit their unusual properties technologically \cite{kiselev:JPDAP:11,fert:NN:13,hellman:RMP:17,garst:JPDAP:17}. Initially identified in a small pocket of the phase diagram of the cubic chiral magnets MnSi, {\fcs}, FeGe, and {\cso}, Lorentz transmission electron microscopy (LTEM) early on revealed a strong increase of the extent of the skyrmion phase with decreasing sample thickness \cite{yu:N:10,yu:NM:11,seki:S:12,tonomura:NL:12,yu:NL:13}. While this demonstrated that nano-scaled systems may be ideally suited to host skyrmions, the mechanisms controlling their formation in samples prepared by established deposition techniques remain unresolved. 

Based on measurements of the magnetization and magneto-transport, interpreted to provide a  topological Hall signal for magnetic fields applied perpendicular to the films, seminal work on epitaxial films of MnSi~\cite{karhu:PRB:10,wilson:PRB:12,li:PRL:13}, 
Mn$_{1-x}$Fe$_{x}$Si~\cite{yokouchi:PRB:14}, 
Fe$_{1-x}$Co$_{x}$Si~\cite{porter:AC:13,sinha:PRB:14} and 
FeGe~\cite{porter:PRB:14,gallagher:PRL:17,ahmed:AC:17} 
claimed the formation of skyrmions. However, the phase diagrams reported for nominally the same films differ between different studies, as well as from those determined by means of LTEM in thin bulk samples. Moreover, alternative mechanisms have been proposed explaining the same data without invoking skyrmions or topological textures ~\cite{monchesky:PRL:14,meynell:PRB:14a}. Subsequent microscopic measurements using polarized neutron reflectometry (PNR)~\cite{karhu:PRB:11,wilson:PRB:14,meynell:PRB:14}, while suggesting modulations perpendicular to the films, failed to resolve this situation convincingly, as PNR requires the assumption of complex scattering profiles. Similar uncertainties arise in studies of epitaxial films using LTEM, either because of parasitic signal interferences in the presence of the substrate~\cite{monchesky:PRL:14,li:PRL:14}, or because in free-standing films removal of the substrate changes the elastic properties of the films \cite{ahmed:AC:17}. Similar uncertainties exist under magnetic fields parallel to the films. Here, magnetization, magneto-transport and PNR have been interpreted to provide putative evidence of complex magnetic phase diagrams supporting the formation of in-plane skyrmions~\cite{karhu:PRB:12,wilson:PRB:12,wilson:PRB:13,yokouchi:JPSJ:15, meynell:PRB:17}. 

Theoretically, it was at first believed, that the increase of the skyrmion phase in thin bulk samples originates from the destabilization of the competing magnetic phases \cite{yu:N:10,yu:NM:11,han:PRB:10}. Recent studies suggest instead that free surfaces favour the formation of skyrmions energetically, driving the formation of skyrmions in surface-dominated systems \cite{meynell:PRB:14,kiselev:PRL:15,zheng:AC:17,wild:SA:17}. Another aspect are interface induced spin-orbit coupling effects, which are dominant for heavy element substrates as used in studies of atomically thin films of Fe and FePd \cite{heinze:NP:2011,romming:S:13}. Last but not least, numerous studies have considered magnetic anisotropies induced by the lattice mismatch with the substrate \cite{butenko:PRB:10,leonov:AC:10,rossler:JPCS:11,wilson:PRB:12,rybakov:PRB:13,wilson:PRB:14}, which, however, are found to decrease rapidly with increasing film thickness. 

To establish the nature of the long-range magnetic order in thin films unambiguously, direct reciprocal space imaging by means of neutron scattering is ideally suited. Meaningful diffraction patterns are already expected for short correlation lengths. Typical features of the magnetic structures considered in epitaxial thin films of B20 compounds can be found in the supplementary information \cite{SOM}. 

A major technical constraint of neutron scattering in thin films is the tiny sample volume. For instance, a recent study on a thick MnSi film using small angle neutron scattering (SANS) in a transmission geometry has demonstrated that even for a measurement time of 20 hours the magnetic signal was barely discernable \cite{ meynell:PRB:17}. In comparison, a clear SANS signal was observed in a stack of 32 FeGe films ($15\times 15 \,{\rm mm^2}$ corresponding to 12\,mg of material) illuminated parallel to the film~\cite{kanazawa:PRB:16}. Unfortunately, this approach requires perfectly identical large films, bearing the additional risk that important details may get averaged out.

In view of these complexities an important point of reference are the differences between thin bulk samples, providing a well-defined setting, and epitaxial layers of comparable thickness. For FeGe on Si(111), where the lattice mismatch is vanishingly small, the SANS on a stack of 32 large films has provided compelling evidence that the magnetic order consists of a helical modulation perpendicular to the film without any evidence for more complex textures, notably skyrmions. The pronounced difference of the magnetic order in epitaxial films and thin bulk samples appears to be well beyond present understanding and underscores the need for further microscopic information on structural differences.

In our study we focus on the properties of epitaxial MnSi films on Si(111) since this allows for comparison with thin bulk samples of MnSi as well as the large body of literature on FeGe. Demonstrating for the first time, that GISANS and OSR can be performed on just a single film rather than a large stack \cite{wiedenmann:phd}, we find a magnetic modulation perpendicular to the MnSi films featuring magnetic phase diagrams under parallel and perpendicular field that are highly reminiscent of those in FeGe on Si(111). Moreover, extended x-ray absorption fine structure  (EXAFS) on the same thick epitaxial MnSi films reveals that the lattice mismatch with the substrate is released within a few atomic layers, such that the films display unstrained lattice parameters, however, with small shifts of the Si positions normal to the film plane \cite{figueroa:PRB:16}. Taken together our results suggest formation of a magnetic modulation perpendicular to the films as a generic property of thick epitaxial films of B20 compounds supported by a substrate, caused by tiny shifts of the non-magnetic atoms.

Epitaxial MnSi(111) films were grown on Si(111) substrates using molecular beam epitaxy following the established recipe for solid-phase epitaxy \cite{karhu:PRB:10,SOM}. For our study we selected three MnSi film thicknesses [$d=\SIA{(390\pm10)}{\AA}, \SIA{(495\pm10)}{\AA}$, and \SIA{(553\pm10)}{\AA}] that are much larger than the length of the magnetic helix of $\sim$180~\AA\ in bulk MnSi, and well above the thickness of $\sim$120~\AA\ where $T_\mathrm{c}$ becomes thickness-independent. Moreover, the induced uniaxial anisotropy is vanishingly small~\cite{karhu:PRB:12,brasse:phd}. The $\sim$550~\AA\ thick sample displayed structural and magnetic properties consistent with data reported in the literature \cite{karhu:PRB:10,karhu:PRB:11,wilson:PRB:12,karhu:PRB:12,li:PRL:13,menzel:JotKPS:13,wilson:PRB:13,wilson:PRB:14,meynell:PRB:14,lancaster:PRB:16,zhang:AA:16}. Samples without capping, Si capping and Cu capping were prepared. The samples with Cu capping  were designed as neutron resonator (waveguide) structure to enhance spin-flip scattering by non-collinear magnetic structures such as skyrmions. However, we did not detect any significant spin-flip scattering in our PNR experiments consistent with the absence of skyrmions. All samples displayed the same magnetic properties apart of small differences of the modulation length of order 10\,\%.

\begin{figure}[t!]
\includegraphics[width=1 \columnwidth]{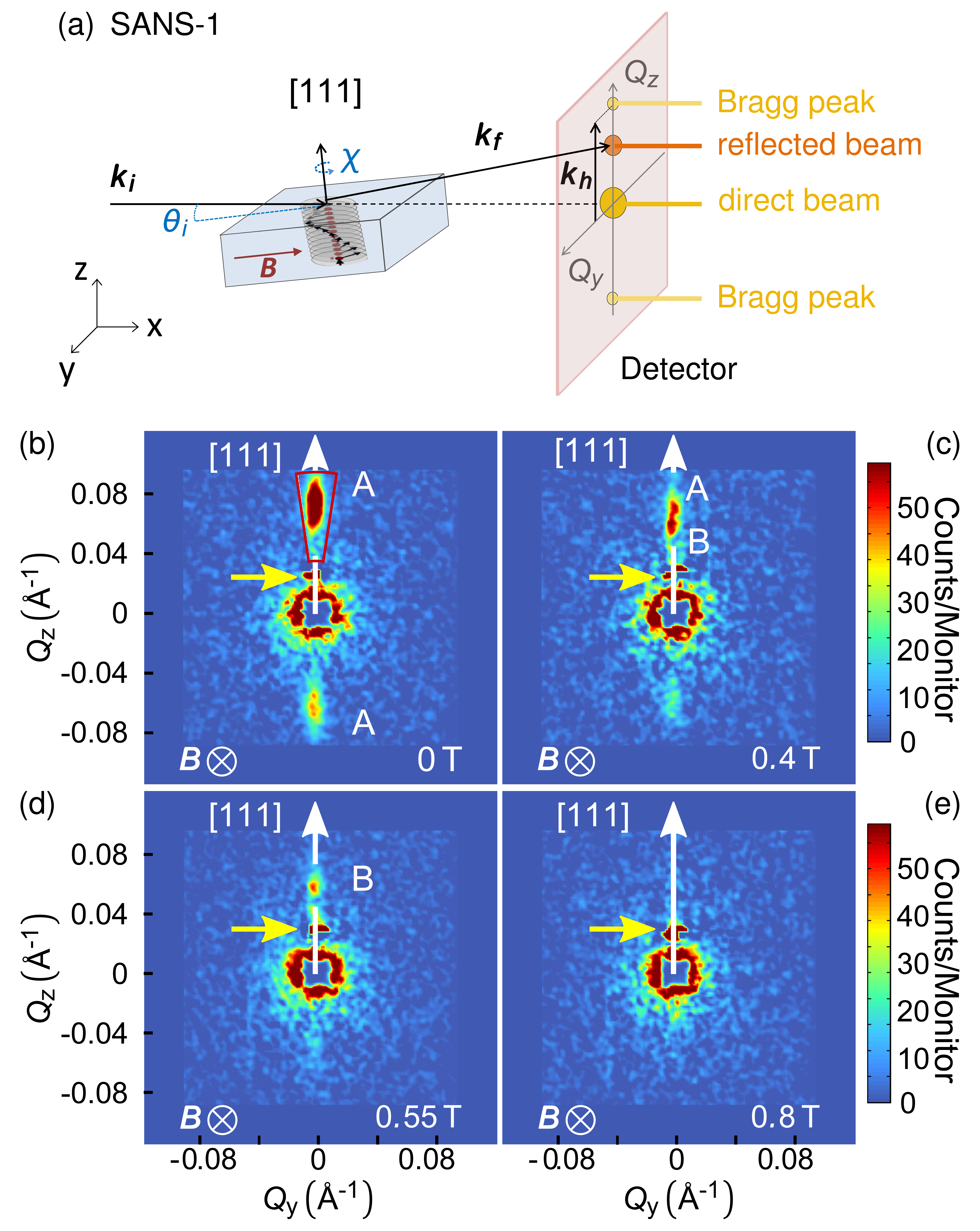}
\caption{Key aspects of the GISANS studies at SANS-1.
(a) Schematic GISANS setup. The incident neutron beam illuminates the MnSi film under grazing incidence.  ${\bm B}$ was applied either parallel or perpendicular (not shown in the figure) to the film.
(b-e) Typical GISANS data for a MnSi thin film (${d=\SI{553}{\AA}}$) at $T= $\SI{15}{K} under increasing magnetic fields which are aligned along the forward direction [cf. panel (a)]. The yellow arrow marks the specularly reflected beam. Peaks labeled A and B arise from magnetic modulations along $Q_z$ corresponding to ${k_h^A=\SI{0.067}{\AA^{-1}}}$ and ${k_h^B=\SI{0.052}{\AA^{-1}}}$, respectively.}
\label{Fig1GISANS} 
\vspace*{-0.5cm}
\end{figure}

Shown in Fig.~\ref{Fig1GISANS}(a) is the GISANS setup used for our study. Measurements were carried out at SANS-1 at the Heinz Maier-Leibnitz Zentrum (MLZ), Munich~\cite{muhlbauer:NIaMiPRSAASDaAE:16}. The sample was illuminated under an incident angle of ${\theta_i = \SI{0.65}{\degree}}$ using neutrons of wavelength ${\lambda = \SIA{(5.5\pm0.5)}{\AA}}$ collimated over \SI{8}{\m} and recorded with a detector \SI{6}{\m} behind the sample. Due to the magnetic mosaicity of the sample and the small scattering angles Bragg peaks for both +$\bm{k}_h$ and -$\bm{k}_h$ could be observed simultaneously. This allowed us to better separate the specular reflection (yellow arrow) from the Bragg peaks. GISANS data was recorded for fields perpendicular as well as parallel to the film. 

Typical data at ${T=\SI{15}{\K}}$ for a MnSi film (${d=\SI{553}{\AA}}$) capped with \SI{310}{\AA} of Cu are shown in Figs.~\ref{Fig1GISANS}(b-e) for a field parallel to the film and the neutron beam [Fig.~\ref{Fig1GISANS}(a)] after subtraction of the background determined at 60~K. The direct beam in the center of the images was partially masked (blue square) to prevent saturation of the detector, yet permitting data analysis even very close to the direct beam. In zero field, a magnetic satellite peak labeled A at ${|Q_z| = \pm\SI{0.067}{\AA^{-1}}}$ is observed, characteristic of a magnetic modulation along \hkl [111] normal to the film plane.  As compared to bulk MnSi the pitch of the modulation is a factor of two smaller. In view of the bulk properties of MnSi we assume the formation of a helical modulation.

Remarkably, with increasing magnetic field the peak sharpens without changing its location and a second peak labelled B emerges above $\sim\SI{0.2}{T}$ at ${|Q_z|\!=\!\SI{0.052}{\AA^{-1}}}$ as shown in Fig.\ \ref{Fig1GISANS}(c) for ${B\!=\!\SI{0.4}{\tesla}}$. When increasing the field, the peaks at A and B sharpen and the peak at A vanishes above $\sim\SI{0.5}{T}$ [Fig.\ \ref{Fig1GISANS}(d)] before the peak at B, which continues to sharpen even further, vanishes above $\sim\SI{0.8}{T}$ [Fig.\ \ref{Fig1GISANS}(e)]. The same behavior was observed for fields parallel to the film, regardless if the fields were parallel or perpendicular to the neutron beam. Similarly, for a magnetic field perpendicular to the film a magnetic modulation perpendicular to the film of unchanged modulation length was observed for all magnetic fields up to the onset of the field-polarised state at $H_{c2}$. Considering instrumental resolution, neither the GISANS data nor any of the other data we recorded provide indications of scattering at finite $Q_y$ or $Q_x$ in the parameter regime explored. Thus the magnetic order in our MnSi film is definitely dominated by a magnetic modulation perpendicular to the film.

\begin{figure}[t!]
\includegraphics[width=1.0\columnwidth]{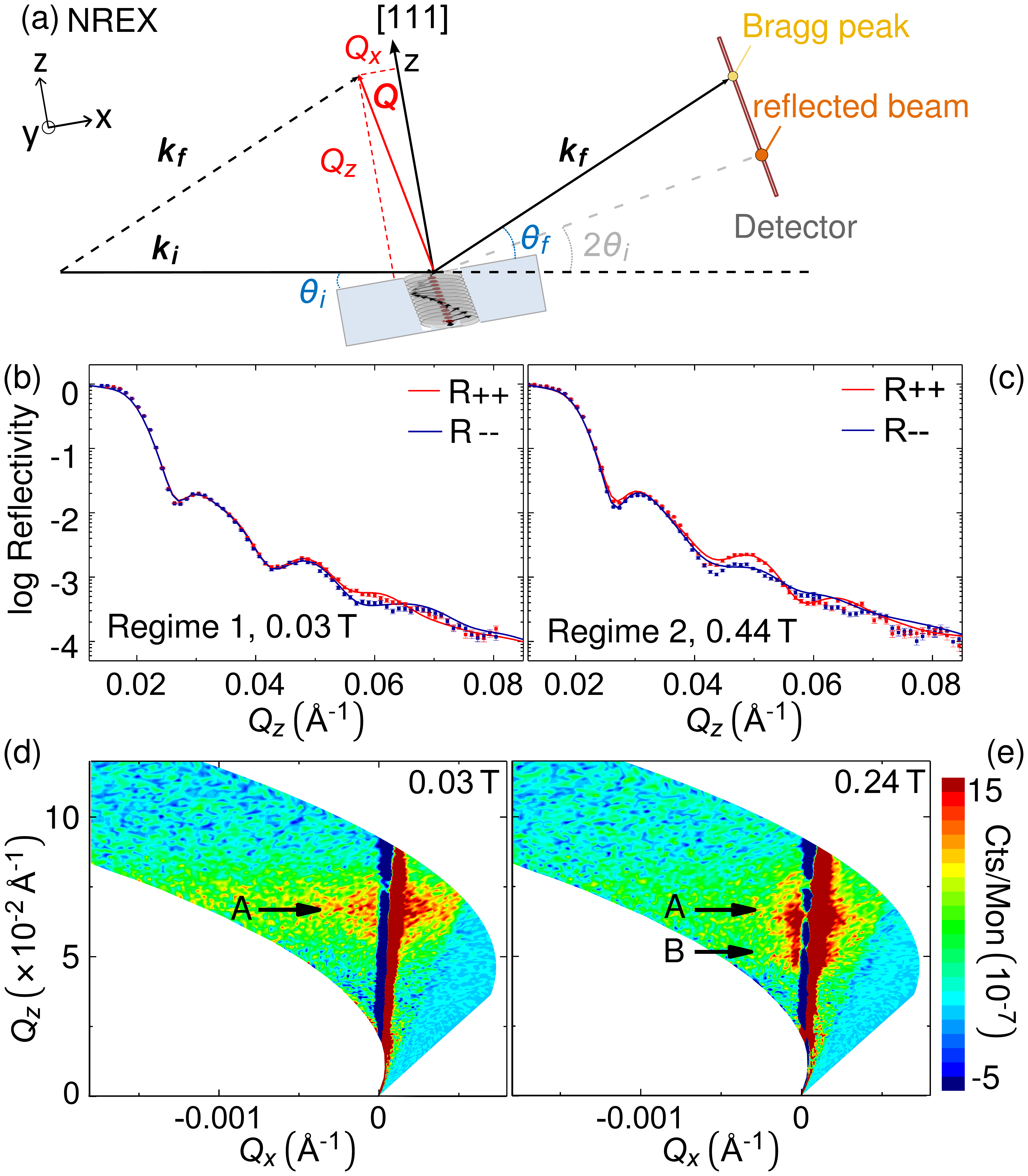}
\caption{
Summary of key data as collected by means of PNR and OSR on a film with a thickness $d = 553$ \AA.
(a) Schematic of the PNR and OSR set-up. The sample was oriented with \hkl[111] perpendicular to the sample surface, and \hkl[11-2] along the $y$-direction. For PNR, the in-plane magnetic field ${\bm B}$ was applied along the $y$-direction, while for OSR $\bm{B}$ was aligned in the forward direction parallel to \hkl[1-10].
(b,c) PNR recorded at $T = 35$~K indicating a magnetic response due to a periodic modulation of the magnetization perpendicular to the film at $B_{[112]} = 30$~mT and $B_{[112]} = 440$~mT corresponding to regimes 1 and 2 (see Figs. 3 and 4), respectively.
(d,e) OSR recorded at $T = 15$\,K for $B_{[112]} = 30$~mT and $B_{[112]} = 240$~mT, where a single modulation and two modulations are seen, respectively.}
\label{Fig2PNR}
\vspace*{-0.5cm}
\end{figure}

To connect the GISANS data with properties reported in the literature ~\cite{karhu:PRB:11, karhu:PRB:12} we performed PNR (${\theta_i =\theta_f}$) and OSR (${\theta_i \neq\theta_f}$) at NREX, MLZ~\cite{Khaydukov:JLRFJ:15}. Using neutrons with a wavelength $\lambda\!=\!\SI{4.31}{\AA}$ and a polarization parallel and antiparallel to the in-plane magnetic field ${\bm B}_y$ [Fig.~\ref{Fig2PNR}(a)] yields an improved $Q_z$ resolution and corroborates the GISANS results. Shown in Fig.~\ref{Fig2PNR}(b,c) is the specular reflectivity of polarized neutrons at ${T=\SI{35}{\K}}$ for the same film investigated in Fig.\,\ref{Fig1GISANS}. 
Now the field is applied along a \hkl [11-2]-direction ($y$-direction) and perpendicular to the incident neutron beam [Fig.~\ref{Fig2PNR}(a)]. 
Data at low fields (${B_{y}=\SI{0.03}{\tesla}}$) may be well accounted for by assuming a magnetic helix with a wavevector ${k_h=\SI{0.067}{\AA^{-1}}}$ parallel to $Q_z$. This result compares with typical PNR at larger fields (${B_{\hkl[11-2]}=\SI{0.44}{\tesla}}$) shown in Fig.~\ref{Fig2PNR}(c), where the reflectivity may be fitted by assuming an anharmonic helix with a smaller wavevector ${k_h=\SI{0.052}{\AA^{-1}}}$ and the magnetic moments pointing predominantly along $\bm {B}_y$. The appearance of helices parallel to $Q_z$ inferred from our PNR data is in excellent agreement with our GISANS results, as well as with the literature \cite{karhu:PRB:11, wilson:PRB:13}.

Shown in Figures~\ref{Fig2PNR}(d) and (e) are typical OSR data of the same MnSi film at ${T=\SI{15}{\K}}$ under an in-plane magnetic field ${\bm B}_{\hkl[1-10]}$ oriented in the forward direction along \hkl [1-10], where a background recorded at 60\,K was subtracted; the largest field accessible was 0.24\,T. The vertical stripes (dark blue and red) for ${Q_x = 0}$ are due to the specular reflection of the neutrons by the sample [cf Figs.~\ref{Fig2PNR}(b,c)]. Here superlattice peaks due to Bragg scattering from the helix, seen in GISANS, are buried by the strong specular signal. The arrows labeled A and B indicate weak magnetic correlations within the plane due to a small amount of roughness leading to the appearance of Bragg sheets at wavevectors ${k_h^A = \SI{0.067}{\AA^{-1}}}$ and ${k_h^B = \SI{0.052}{\AA^{-1}}}$. The position of these peaks corresponds to the helix vectors inferred from PNR as well as from GISANS. The weak intensity of the Bragg sheets is an indication for (i) the excellent quality of the samples and (ii) the excellent alignment of the helices perpendicular to the film even in large  fields perpendicular to $\bm {k}_h$. 

We note that PNR measurements in a MnSi film with a similar thickness of \SI{267}{\AA}, reported in the literature \cite{karhu:PRB:12,wilson:PRB:14}, were satisfactorily fitted with a sinusoidal magnetization profile and a modulus of the helix vector of \SI{0.045}{\AA^{-1}} in zero field and a distorted sinusoidal shape together with a different periodicity for in-plane magnetic fields of 0.2\,T and 0.4\,T. This discrepancy with our data may be attributed to a reduced sample size, but reflects also the uncertainties when interpreting PNR data.

\begin{figure}[t!]
\includegraphics[width=1.0\columnwidth]{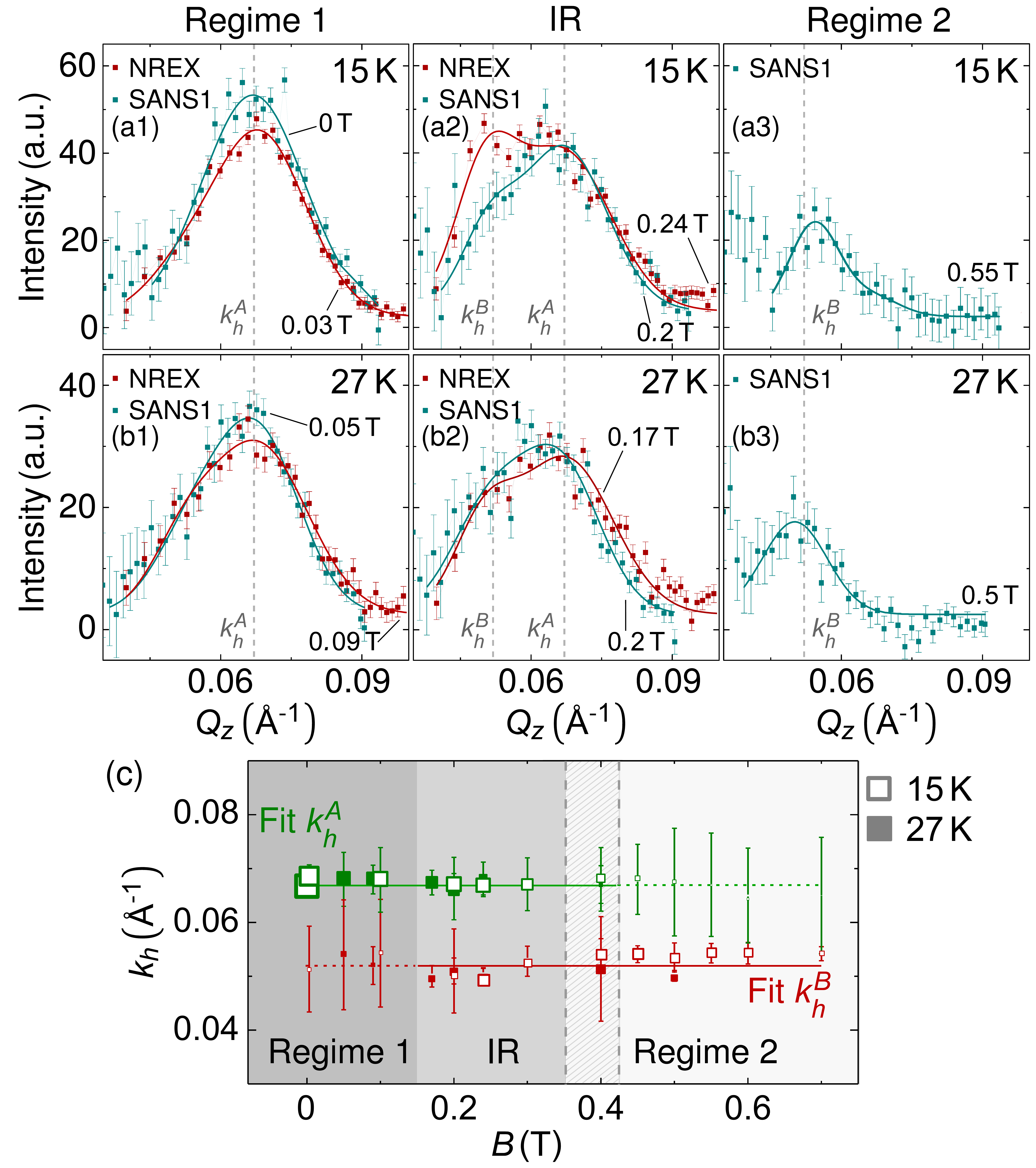}
\caption{Comparison of the $Q_z$-dependence observed in GISANS and OSR for various fields and temperatures. (a1,a2,a3) Field-dependence at 15\,K, and (b1,b2,b3) at 27\,K. The solid lines represent Gaussian fits. (c) Field dependence of the position of the magnetic satellite peaks A and B. The regimes 1 and 2 are highlighted in dark gray and white, respectively. Dotted vertical lines depict the transition from the intermediate regime to regime 2.}
\label{Fig3VglNrexSANS}
\vspace*{-0.5cm}
\end{figure}

A comparison of the $Q_z$-dependence observed in GISANS and OSR is shown in Figure~\ref{Fig3VglNrexSANS}. For this comparison the GISANS data was integrated radially within the sector indicated by the red lines shown in Fig.~\ref{Fig1GISANS}(b). The data was scaled with a single normalisation constant for the comparison with the OSR. The $Q_z$-dependence of the OSR signal was obtained by integrating the intensity on the negative side of the specular signal for momentum transfers ${Q_x \le\SI{-3e-6}{\AA^{-1}}}$ [see Figs.~\ref{Fig2PNR}(d,e)]. All data (Fig.~\ref{Fig3VglNrexSANS}) displays a single peak at low (a1,b1) and high fields (a3,b3), respectively, and two peaks at intermediate fields (a2,b2), although the measurements have been conducted under very different resolution conditions and for two different spatial orientations with respect to the in-plane magnetic field. As no diffraction peaks or intensities are observed in the $Q_x$--$Q_y$ plane, the modulations evolve along $Q_z$ only. Even though the GISANS, OSR and PNR signals cannot be tracked all the way up to $B_{c2}$ and $T_c$, the accessible $T$ and $B$ ranges are sufficiently large to note that we do not find any evidence supporting earlier claims of in-plane skyrmions \cite{wilson:PRB:12, meynell:PRB:17}.

A similar reduction of the helical modulation length under magnetic field parallel to the film was previously proposed in a related study of a film with $d=267\,{\rm \AA}$ using PNR, magneto-transport and magnetometry \cite{wilson:PRB:13}. Our data are also consistent with an unwinding in discrete steps, also reported in Ref.\,\cite{wilson:PRB:13}. This behaviour agrees with the interplay of a perpendicular magnetic anisotropy, favouring an out-of-plane propagation, and the Zeeman energy under applied fields. It is interesting to note, that the magnetic anisotropy constant inferred in our films from torque magnetometry is vanishingly small \cite{brasse:phd} and consistent with estimates based on the magnetisation reported in the literature \cite{karhu:PRB:12}. This result suggests a more subtle origin of the propagation direction other than strain-induced magnetic anisotropies. 

\begin{figure}[t!]
\includegraphics[width=1\columnwidth]{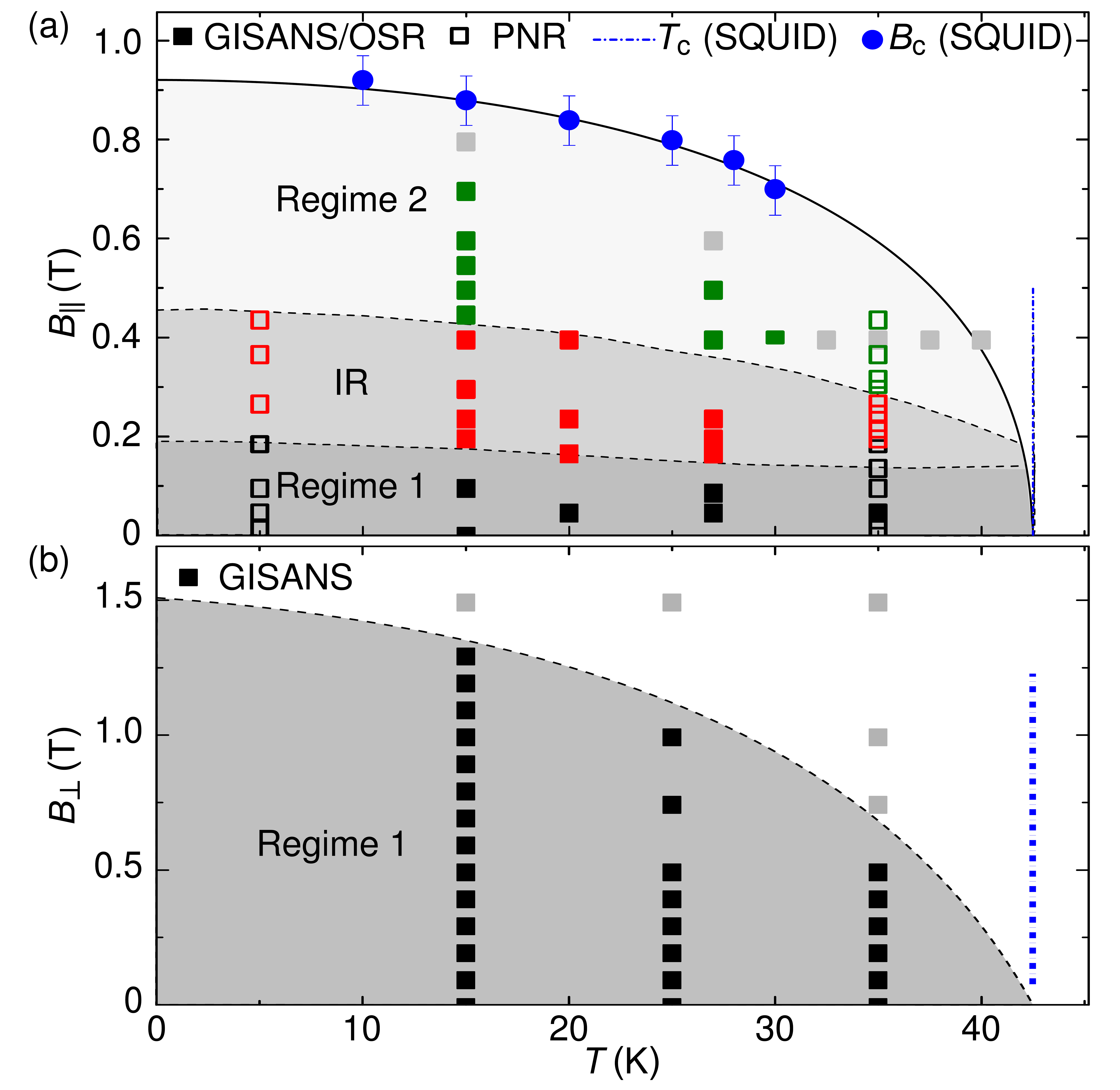}
\caption{$(B,T)$--phase diagrams for an epitaxial MnSi film ($d=553\,{\rm\AA}$) as inferred from GISANS, PNR, OSR and SQUID magnetometry. No magnetic signal was observed with GISANS at the gray-shaded positions.
(a) Phase diagram under magnetic field parallel to the film. Two regimes may be distinguished at low fields and high fields with distinct helical modulations, $k_h^A$ and $k_h^B$, respectively. At intermediate fields both modulations coexist.
(b) Phase diagram under magnetic field perpendicular to the film. A single helical modulation $k_h^A$ perpendicular to the film is observed.}
\label{figure4}
\vspace*{-0.5cm}
\end{figure}

Combining our neutron data with SQUID magnetometry on the same MnSi films the magnetic phase diagram shown in Fig.~\ref{figure4} was constructed. The PNR data (open symbols) and GISANS/OSR data (filled circles) are fully consistent with each other. The phase boundaries between regime 2 and the field-aligned ferromagnetic phase and $T_\mathrm{c}$ were inferred from the SQUID data. The upper critical field $B_{c2} = 0.93\,{\rm T}$ (blue dots), as well as $T_\mathrm{c} = 42.5\,{\rm K}$, are significantly higher than the corresponding values in bulk MnSi, in good agreement with the literature \cite{li:PRL:13,karhu:PRB:12, wilson:PRB:12}.

The magnetic order and the phase diagrams we observe are in striking similarity with the magnetic phase diagram observed in a SANS study of a stack of 32 FeGe thin films. For the FeGe films the lattice mismatch is tiny, rendering simple strain-induced magnetic anisotropies highly unlikely. For our MnSi films, EXAFS \cite{figueroa:PRB:16} establishes, that the lattice strain is released within a few atomic layers, while the main part of the film is unstrained apart from tiny out-of-plane shifts of the positions of the Si atoms. These shifts give way for an unexpected new mechanism controlling the appearance of an out-of-plane modulation in epitaxial MnSi films.

In summary, we present the first reciprocal space mapping of the magnetic order in single thick epitaxial MnSi films on Si(111) demonstrating the potential of GISANS and OSR, where we find a magnetic phase diagram in striking similarity with thick epitaxial FeGe films. Combining these results with EXAFS and the established properties of thin bulk samples, we conclude that the magnetic order we observe is generic, providing an entry port for unravelling the rather conflicting information on thin films in the literature as concerns the precise mechanisms that allow to tailor skyrmions in nanosystems by means of established deposition techniques.

\acknowledgments
We gratefully acknowledge technical support by the team of FRM II as well as Franz Tralmer at NREX. We also wish to thank A. Bauer, M. Brasse, D. Grundler, C. Schnarr and M. Wilde for discussions and support. Financial support through the collaborative research center TRR80 of the German Science Foundation (DFG) projects E4 and F7, as well as ERC Advanced Grant 291079 (TOPFIT) are gratefully acknowledged. BW, AC, MH and TA acknowledge financial support through the TUM graduate school. TH and SLZ acknowledge the Semiconductor Research Corporation (SRC) and EPSRC (UK).

\vspace*{-0.5cm}

%

\end{document}


\newcommand{\todo}[1]{\textbf{\textsc{\textcolor{red}{(TODO: #1)}}}}
\newcommand{\OLD}[1]{{\tiny {\bf old:} #1 }}
\newcommand{\NEW}[1]{{ \it #1 }}
\renewcommand{\vec}[1]{\boldsymbol{#1}}
\newcommand{\w}{\omega}

\newcommand{\fcs}{Fe$_{1-x}$Co$_{x}$Si}
\newcommand{\mfs}{Mn$_{1-x}$Fe$_{x}$Si}
\newcommand{\mcs}{Mn$_{1-x}$Co$_{x}$Si}
\newcommand{\cso}{Cu$_{2}$OSeO$_{3}$}

\newcommand{\rxx}{$\rho_{\rm xx}$}
\newcommand{\rxy}{$\rho_{\rm xy}$}
\newcommand{\rxyt}{$\rho_{\rm xy}^{\rm top}$}
\newcommand{\Drxyt}{$\Delta\rho_{\rm xy}^{\rm top}$}
\newcommand{\Sxy}{$\sigma_{\rm xy}$}
\newcommand{\Sxya}{$\sigma_{\rm xy}^A$}

\newcommand{\bco}{$B_{\rm c1}$}
\newcommand{\bct}{$B_{\rm c2}$}
\newcommand{\bao}{$B_{\rm A1}$}
\newcommand{\bat}{$B_{\rm A2}$}
\newcommand{\beff}{$B^{\rm eff}$}

\newcommand{\btr}{$B^{\rm tr}$}

\newcommand{\tc}{$T_{\rm c}$}
\newcommand{\ttr}{$T_{\rm tr}$}

\newcommand{\mb}{$\mu_0\,M/B$}
\newcommand{\dmdb}{$\mu_0\,\mathrm{d}M/\mathrm{d}B$}
\newcommand{\ddmddb}{$\mathrm{\mu_0\Delta}M/\mathrm{\Delta}B$}
\newcommand{\cac}{$\chi_{\rm ac}$}
\newcommand{\rechi}{${\rm Re}\,\chi_{\rm ac}$}
\newcommand{\imchi}{${\rm Im}\,\chi_{\rm ac}$}

\newcommand{\ozz}{$\langle100\rangle$}
\newcommand{\ooz}{$\langle110\rangle$}
\newcommand{\ooo}{$\langle111\rangle$}
\newcommand{\too}{$\langle211\rangle$}

\makeatletter
\renewcommand{\thefigure}{S\@arabic\c@figure}
\makeatother

\renewcommand{\floatpagefraction}{0.5}
\setcounter{secnumdepth}{2}

\title{Supplementary Material for:\\ 
Reciprocal space mapping of magnetic order in\\ 
thick epitaxial MnSi films}

\author{B. Wiedemann}
\address{Physik Department, Technische Universit\"at M\"unchen, James-Franck-Strasse 1, 85748 Garching, Germany}

\author{A. Chacon}
\address{Physik Department, Technische Universit\"at M\"unchen, James-Franck-Strasse 1, 85748 Garching, Germany}

\author{S. L. Zhang}
\address{Clarendon Laboratory, Department of Physics, University of Oxford, Parks Road, Oxford, OX1~3PU, UK}

\author{Y. Khaydukov} 
\address{Max-Planck-Institut f\"ur Festk\"orperforschung, Heisenbergstrasse 1, 70569 Stuttgart, Germany}
\address{Max Planck Society, Outstation at FRM-II, D-85748, Garching, Germany}

\author{T. Hesjedal}
\address{Clarendon Laboratory, Department of Physics, University of Oxford, Parks Road, Oxford, OX1~3PU, UK}

\author{O. Soltwedel} 
\address{Physik Department, Technische Universit\"at M\"unchen, James-Franck-Strasse 1, 85748 Garching, Germany}
\address{Max-Planck-Institut f\"ur Festk\"orperforschung, Heisenbergstrasse 1, 70569 Stuttgart, Germany}
\address{Max Planck Society, Outstation at FRM-II, D-85748, Garching, Germany}

\author{T. Keller} 
\address{Max-Planck-Institut f\"ur Festk\"orperforschung, Heisenbergstrasse 1, 70569 Stuttgart, Germany}
\address{Max Planck Society, Outstation at FRM-II, D-85748, Garching, Germany}

\author{S. M\"uhlbauer} 
\address{Forschungsneutronenquelle Heinz Maier Leibnitz (FRMII), Technische Universit\"at M\"unchen, 85748 Garching, Germany}

\author{T. Adams}
\address{Physik Department, Technische Universit\"at M\"unchen, James-Franck-Strasse 1, 85748 Garching, Germany}

\author{M. Halder}
\address{Physik Department, Technische Universit\"at M\"unchen, James-Franck-Strasse 1, 85748 Garching, Germany}

\author{C. Pfleiderer}
\address{Physik Department, Technische Universit\"at M\"unchen, James-Franck-Strasse 1, 85748 Garching, Germany}

\author{P. B\"oni}
\address{Physik Department, Technische Universit\"at M\"unchen, James-Franck-Strasse 1, 85748 Garching, Germany}

\date{\today}

\begin{abstract}
Supplementary information is given on the scattering patterns expected for different forms of magnetic order considered in epitaxial films of B20 compounds. Additional information concerning the experimental methods is summarized, and additional experimental data reported that complement the results shown in the main text. 
\end{abstract}


\vskip2pc

\maketitle


\section{Fourier Transformations of Different Magnetic Orderings
\label{Fourier}}

Magnetic order in thin epitaxial films of B20 compounds considered in the literature comprise (i) helical modulations out-of-plane, (ii) helical modulations in plane , (ii) skyrmion lattices out-of-plane, and (iv) skyrmion lattices in-plane. In order to support the discussion of the experimental data we present in the following the Fourier transform of these forms of magnetic order, convoluted with a Gaussian distribution function to simulate the effects of instrument resolution. The calculated diffraction patterns may be compared directly with the experimental data shown in the main text.

The helical magnetic order was modelled using the conventional representation:
\begin{equation}
\bm{m}(\bm{r}) = \left(\begin{matrix} \cos (\bm{k}_i\cdot\bm{r}\frac{2\pi}{\lambda})\\ \sin(\bm{k}_i\cdot\bm{r}\frac{2\pi}{\lambda}) \\ 0\end{matrix}\right)
\end{equation}
where the propagation vector is aligned with the $z$-axis. For the skyrmion lattices the structure was modelled using:
\begin{equation}
 \bm{m}(\bm{r})=\sum_{i=1,2,3}\left(\left\{\begin{matrix} 0\\0\\-1\end{matrix}\right\}\cos\left(\bm{k}_i\cdot\bm r \frac{2\pi}{\lambda}\right)-\left(\left\{\begin{matrix}0\\0\\-1\end{matrix}\right\}\times \bm{k}_i\right)\sin\left(\bm{k}_i\cdot\bm{r}\frac{2\pi}{\lambda}\right)\right),
\end{equation}
where 
\begin{equation}
\bm{k}_i = \left\{\begin{matrix} \cos(\alpha_i)\sin(\pi/2)\\\sin(\alpha_i)\sin(\pi/2)\\\cos(\pi/2)\end{matrix}\right\}
\end{equation}
and $\alpha_{1,2,3} =\SI{0}{\degree},\SI{120}{\degree}, \SI{240}{\degree}$ represent the angles under which the three helices  in the $x$-$y$ plane are superposed.

In-plane skyrmions or in-plane propagation of helical order were addressed by rotating the grid on which the magnetic structure was calculated. The nomenclature of the results is based on a sample-centered coordinate system, where the out-of-plane direction is parallel to $\hat{z}$.

For our simulations, we assumed a film of MnSi with a thickness of 500 \AA\ and a wavelength of the modulation given by the bulk value $\lambda = \SI{180}{\AA}$. In comparison, the experimentally determined pitch in our thin films was $\sim$$\SI{100}{\AA}$. However, the qualitative appearance of the diffraction patterns is not affected by the difference in $\lambda$. To avoid spectral leakage a three-dimensional Blackman window function was applied on the magnetization distribution before the Fourier transformation was calculated.

\begin{figure}
	\includegraphics[width=1\columnwidth]{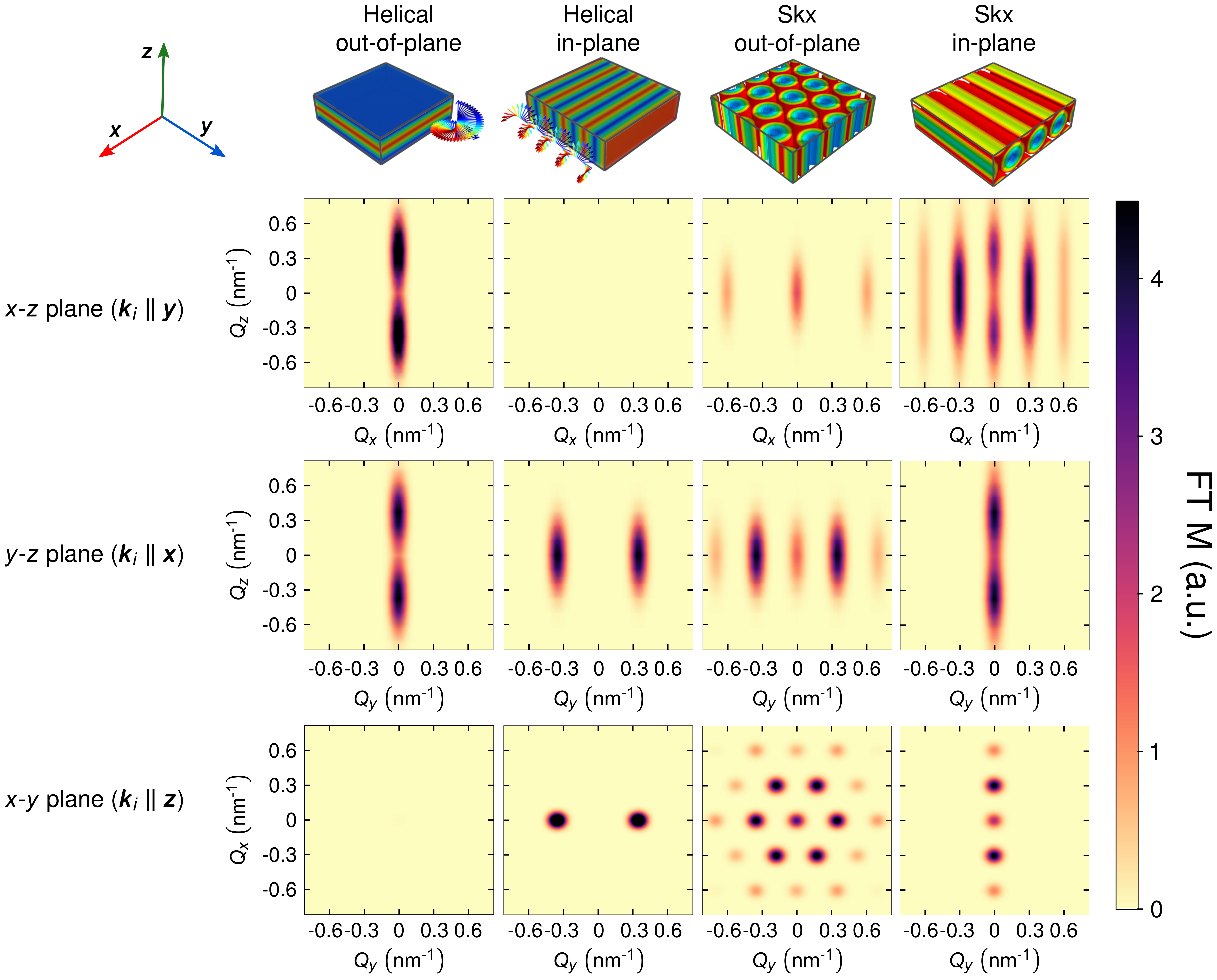}
	\caption{Fourier transforms of the magnetization density of helical and skyrmionic order of a MnSi film that is assumed to have a thickness of 500 \AA\ and a pitch $\lambda = 180$ \AA. The top figures show a sketch of the relevant magnetization distributions. Diffraction peaks along $Q_z$ are only observed for the out-of-plane helix and in-plane skyrmion lattice (see $2^{nd}$ and $3^{rd}$ row). The latter, however, leads to diffraction peaks along the $Q_x$ direction as shown in the $3^{rd}$ and $4^{th}$ row, which have neither been observed experimentally not reported in the literature for epitaxial MnSi films on Si(111) so far (see main text).} 
	\label{fig:Masterplot}
\end{figure}

The results of the calculations are summarized in Fig.\ \ref{fig:Masterplot}. The magnetic structures studied are represented in the top row, while the three main orientations in reciprocal space are shown from top to bottom, namely:
\begin{itemize}

\item $x$-$z$ plane analogous to a GISANS experiment.
\item $y$-$z$ plane analogous to a GISANS experiment after rotating the sample by \SI{90}{\degree} around the $Q_z$-direction.
\item $x$-$y$ plane representing a typical SANS experiment in transmission geometry.
\end{itemize}

The out-of-plane skyrmion lattice and the in-plane helical propagation do not show any signal along $Q_z$, and are thus both incompatible with our GISANS data.
The in-plane skyrmion lattice would show a small signal along $Q_z$. Rotating the sample by \SI{90}{\degree} results in a large signal along either $Q_x$ or $Q_y$. For an in-plane skyrmion lattice, predicted theoretically to form parallel to a field applied in-plane, a scattering intensity along $Q_x$ is expected. Such an easy-to-spot signal was not observed in our measurements. As all orientations were studied experimentally, the only magnetic structure compatible with our results is an out-of-plane helically ordered state.

\newpage

\section{\label{Methods}Experimental Methods}
\subsection{Sample Preparation}

Epitaxial MnSi(111) thin films were grown on Si$(111)$ substrates measuring 20 $\times$ 24~mm$^2$ using molecular beam epitaxy (MBE). The Balzers MBE system has a base pressure of $5 \times 10^{-10}$~mbar and is equipped with electron beam evaporators and effusion cells. Flux control is achieved via cross-beam mass spectrometry. Prior to loading, the Si wafers were first degreased, followed by etching in hydrofluoric acid and oxidation by H$_2$O$_2$. Annealing at 990$^{\circ}$C, and growing a Si buffer layer, leads to the $7 \times 7$ reconstruction, as confirmed by reflection high energy electron diffraction (RHEED).

The sample is then cooled down to room-temperature and $\sim$3 monolayers of Mn were deposited before they are reacted with the Si surface at an elevated temperature of $\sim$$250-300$~$^\circ$C. One monolayer (ML) corresponds to $7.82 \times 10^{14}$ atoms/cm$^2$. This leads to the formation of an epitaxial MnSi seed layer. The MnSi layer has a $(\sqrt{30} \times \sqrt{30})$R30$^\circ$ structure, as determined by RHEED. The subsequent MnSi growth is by the stoichiometric supply of Mn and Si. The growth proceeds up to a thickness of roughly 500\,\AA\ without any signs of the formation of a secondary phase.

The MnSi film thickness as determined by x-ray reflectometry (XRR) (see Subsec.\ \ref{subsec:XRR-XRD}, Fig.\ \ref{fig:XRR+XRD} below) is (553 $\pm$ 10)~\AA. The epitaxial relationship is as follows: Si(111) $\parallel$ MnSi(111) and Si$[11\bar{2}]$ $\parallel$ MnSi$[1\bar{1}0]$. A Cu capping layer with a thickness $d_\textrm{Cu} \approx 320$~\AA\ was deposited onto one sample (SI121) to enhance the neutron scattering cross-section within the MnSi layer and to protect the MnSi layer from oxidation.

\begin{figure}
\centering
\includegraphics[width=0.65\columnwidth]{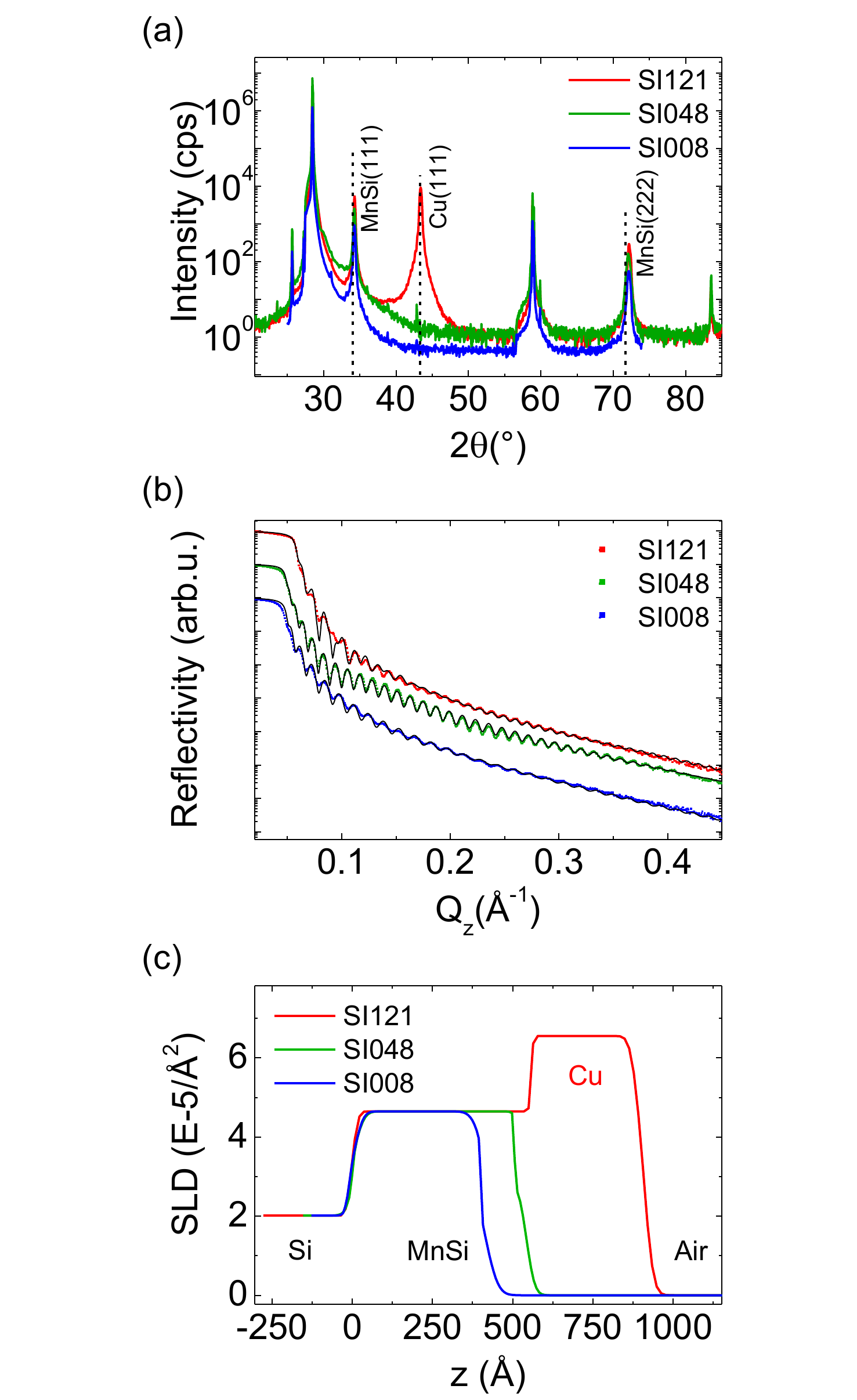}
\caption{(a) Diffraction pattern for various MnSi thin film samples. The out-of-plane lattice parameter of MnSi is reduced by less than 1\% when compared with the bulk value. One of the samples (SI121) was covered with a Cu layer. The unlabeled peaks are from the Si$(111)$ substrate. (b) Reflectivity data for the three MnSi thin film samples. The numerous Kiessig fringes indicate well-defined interfaces. The data were fitted using the Parratt algorithm (black lines). (c) Depth profiles of the scattering length density (SLD) as obtained from the fits. The MnSi films have thicknesses of $d_\textrm{MnSi} = 390,\,495$, and 553\,\AA, respectively.
}
\label{fig:XRR+XRD}
\end{figure}

\newpage
\section{\label{Data}Experimental Data}
\subsection{\label{subsec:XRR-XRD}X-Ray Diffraction and Reflectivity}

For the determination of the out-of-plane lattice parameters, the perfection of the interfaces, and the thickness of the epitaxial films, x-ray diffraction (XRD) and XRR scans were carried out on a D5000 diffractometer using Cu K$_{\alpha 1}$ radiation. The momentum transfer $\bf Q$ was always oriented perpendicular to the film surface. The reflectivity data was fitted using the Parratt32 algorithm \cite{Parratt32,Parratt1954} from which the scattering length density (SLD) of the sample can be determined.

XRD patterns from the MnSi films are shown in Fig.\ \ref{fig:XRR+XRD}(a). The unlabeled peaks indicate reflections from the Si$(111)$ substrate, and the MnSi (and Cu) peaks are indicated. The high intensity and the narrow linewidth of the MnSi $(111)$ and $(222)$ peaks confirm that the MnSi has grown epitaxially in the (111) orientation, confirming the RHEED measurements. Fitting the peak positions yields an interplanar distance $d_{111}$ which is less than 1\% smaller than the bulk value.

The XRR curves, shown in Fig.\ \ref{fig:XRR+XRD}(b), yield numerous Kiessig fringes, up to $Q_z =0.4$\,\AA, indicating well-defined interfaces. From the fits (solid lines) using the Parratt algorithm~\cite{Parratt32} layer thicknesses of $d = (390 \pm 10)$\,\AA, $(495 \pm 10)$\,\AA, and $(553 \pm 10)$\,\AA, are extracted for the samples SI008, SI1048, and SI121, respectively. For the MnSi/Si interface, a roughness between 22 and 30\,\AA\ is obtained.

Finally, the scattering length density profile was extracted which is shown in Fig.\ \ref{fig:XRR+XRD}(c). The Si/MnSi interface is located at $z = 0$. Obviously, the profiles show similar characteristics demonstrating that MnSi samples can be grown reproducibly. Sample SI121 is covered with a Cu layer to both enhance the neutron cross-section and to protect the surface from oxidation.

\subsection{Magnetization Measurements}

\begin{figure}
	\includegraphics[width=0.5\columnwidth]{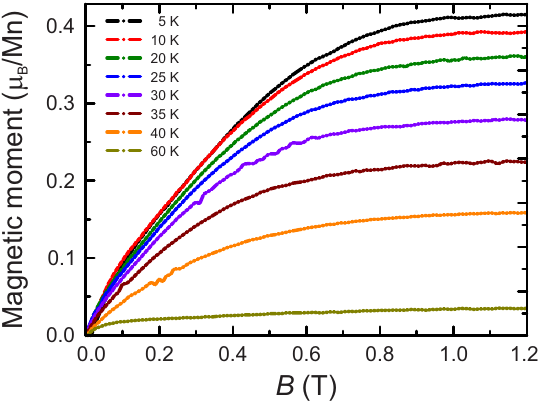}
	\caption{Magnetization as a function of magnetic field at selected temperatures for a 50\,nm thick MnSi thin film under field applied out-of-plane, i.e., along the MnSi[111] direction.}
	\label{fig:MagProp}
\end{figure}

The magnetic properties of the films were also determined with superconducting quantum interference device (SQUID) magnetometry using a Quantum Design SQUID with a vibrating sample magnetometer module. Magnetization measurements were performed as a function of temperature, for the field applied in-plane for a 50-nm-thick MnSi film on Si(111). The results shown in Fig.\ \ref{fig:MagProp} were obtained by field-cooling from 300~K to the indicated temperature in an applied field of 2~T. 
The diamagnetic background stemming from the Si substrate, along with contributions from unsaturated Mn moments at high field, were subtracted. The saturation magnetization, at 5~K, is 0.41(3)~$\mu_\textrm{B}$/Mn, consistent with the value obtained for bulk MnSi \cite{Lee2007,Bloch1957}.
The shape of the curves is consistent with previous studies on epitaxial MnSi films measured in an in-plane applied field. 
From $M(T)$ measurements, a transition temperature of $T_\textrm{c} = 42.3(2)$~K was determined, consistent with the value reported in the literature for epitaxial MnSi films of the same thickness \cite{Karhu11,Karhu12,Li13,Wilson13-helicoidal,Wilson14}.

\subsection{Complementary Neutron Scattering Data}

\begin{figure}
	\includegraphics[width=1\columnwidth]{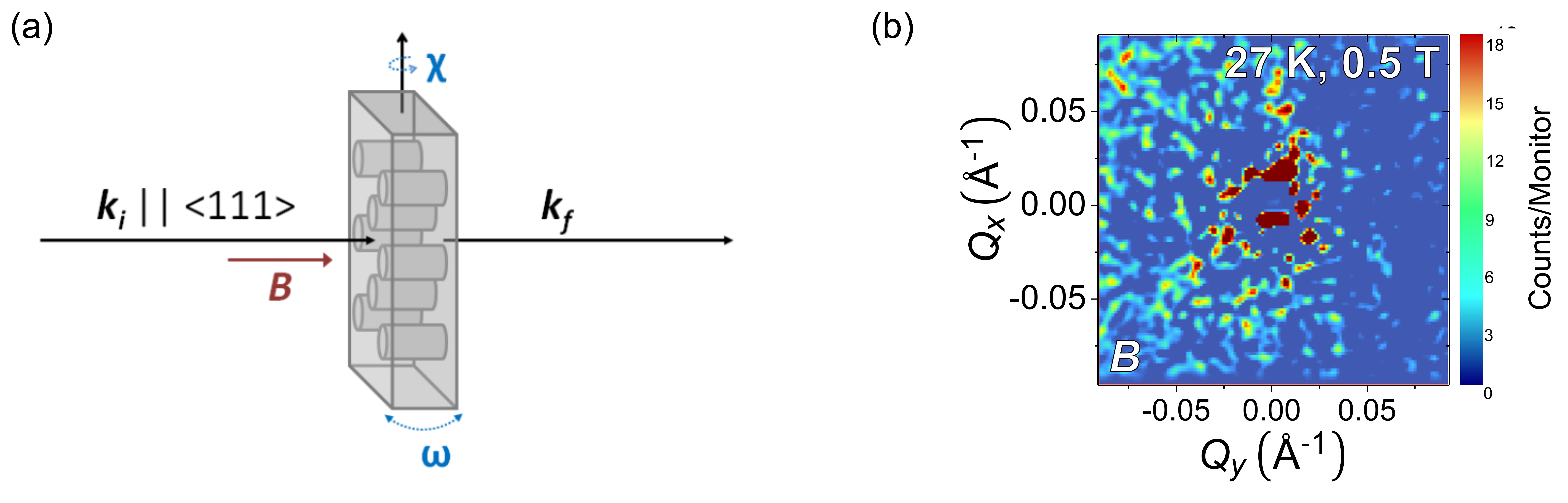}
	\caption{(a) SANS experimental setup in transmission geometry with the magnetic field applied parallel to the neutron beam. (b) Typical scattering pattern obtained in such a geometry, showing no magnetic signal.}
	\label{fig:SANSData}
\end{figure}

In the case of an out-of-plane skyrmion lattice, as proposed in Ref.\ \cite{Li13}, a small angle neutron scattering experiment in the conventional transmission geometry (cf.\ Fig.\ \ref{fig:SANSData}) would yield the six-fold symmetric pattern associated with its hexagonal lattice. We performed such experiments at the instrument SANS-1 at MLZ \cite{Muhlbauer16}. 
The experimental setup is shown in Fig.\ \ref{fig:SANSData}(a). The magnetic field is applied parallel to the neutron beam as in Ref.\ \cite{Li13}. Figure \ref{fig:SANSData}(b) shows typical data gathered in such a scattering geometry. We found no magnetic signal, indicating the absence of magnetic correlations (for example skyrmions) within the plane of the sample. 

\begin{figure}
	\includegraphics[width=1\columnwidth]{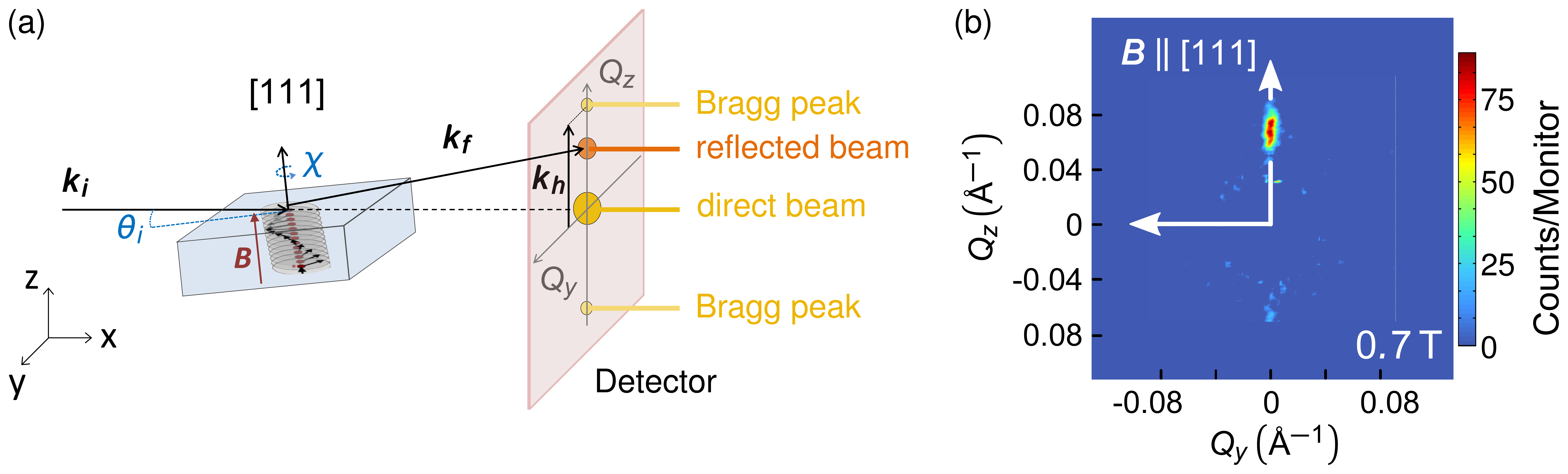}
	\caption{(a) GISANS experimental setup with the magnetic field applied perpendicular to the film. (b) A typical scattering pattern obtained in such a geometry shows a helical propagation perpendicular to the film.}
	\label{fig:GISANSPara}
\end{figure}

Finally, we prove that the magnetic modulations propagate  perpendicular to the film independent of the direction of the applied magnetic field $\bf B$. To prove this statement we have applied $\bf B$ perpendicular to the wavevector ${\bf k}_i$ of the incident neutrons. In a first configuration we applied $\bf B$ perpendicular to the film as shown in Fig.\ \ref{fig:GISANSPara}(b). The data in Fig.\ \ref{fig:GISANSPara}(b) shows that the propagation vector is observed along $Q_z$. With increasing field, a smooth transition to a field-polarized state is observed.

\begin{figure}
	\includegraphics[width=1\columnwidth]{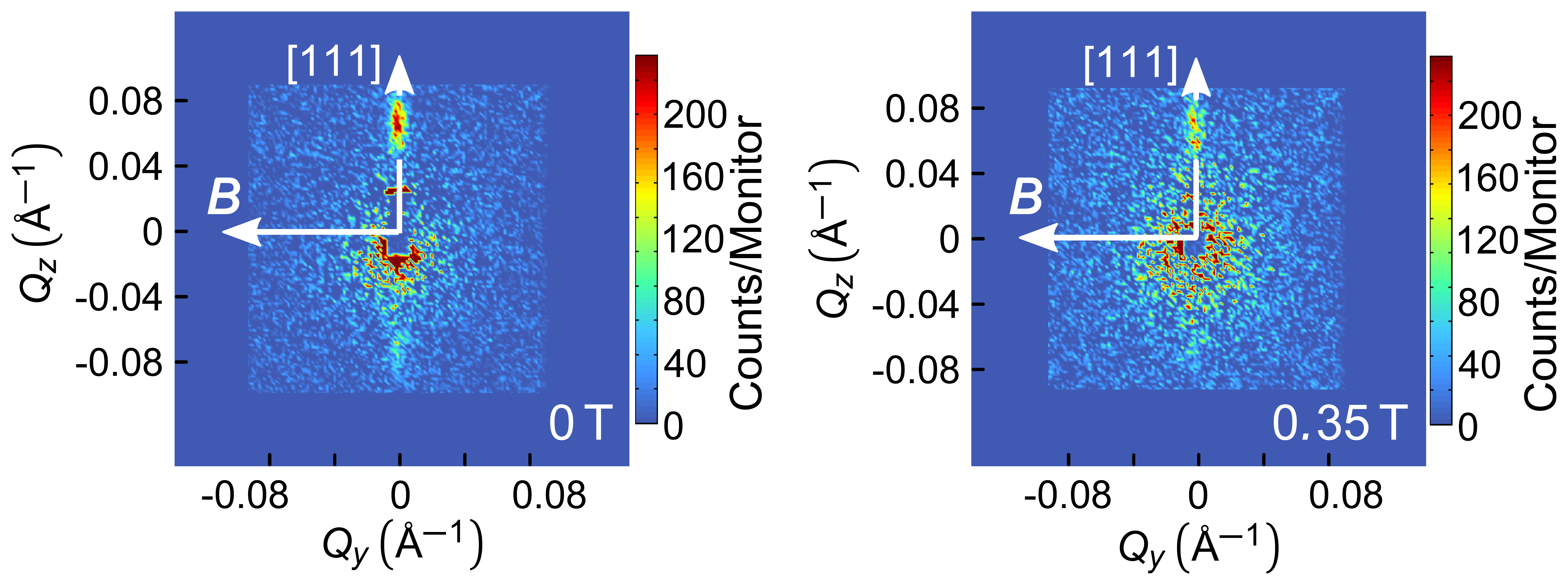}
	\caption{(a) Scattering pattern obtained at $B=\SI{0}{\tesla}$. As expected, only a modulation perpendicular to the film is observed. (b) Scattering pattern obtained in the intermediate field regime. Here, a splitting of the magnetic signal is observed, however, the propagation vectors remain perpendicular to the thin film.}
	\label{fig:GISANSPerp}
\end{figure}

In a second configuration we applied $\bf B$ within the plane perpendicular to ${\bf k}_i$. Typical data in zero field and at an intermediate field $B=\SI{0.35}{\milli\tesla}$ is shown in Fig.\ \ref{fig:GISANSPerp}. The results reproduce the results obtained for the magnetic field applied parallel to the neutron beam (${\bf B} \parallel {\bf k}_i$). The only modulation observed is perpendicular to the film, i.e. along $Q_z$. The expected splitting of the magnetic peaks at intermediate fields is reproduced.

\clearpage
